\documentclass[aps,preprint,prd]{revtex4}
\usepackage{amssymb}
\usepackage{amsmath}
\usepackage{graphicx}

\begin{document}

\title{Triple seesaw mechanism.}

\author{D. Cogollo}
\affiliation{{Departamento de
F\'{\i}sica, Universidade Federal de Campina Grande, Caixa Postal 10071, 58109-970,
Campina Grande, Para\'{i}ba, BraSil }}
\author{H. Diniz}
\affiliation{{ Departamento de
F\'{\i}sica, Universidade Federal da Para\'\i ba, Caixa Postal 5008, 58051-970,
Jo\~ao Pessoa, PB, Brasil}}

\author{C.A. de S. Pires}
\email{cpires@fisica.ufpb.br}
\affiliation{{ Departamento de
F\'{\i}sica, Universidade Federal da Para\'\i ba, Caixa Postal 5008, 58051-970,
Jo\~ao Pessoa, PB, Brasil}}

\date{\today}

\begin{abstract}
%%%%%%%%%%%%%%%%%%%%%%%%%%%%%%%%%%%%%
 On  fitting the type II seesaw mechanism into the type I seesaw mechanism, we obtain a formula to the neutrino masses which get suppressed by high-scale $M^3$ in its denominator. As  a result, light neutrinos are naturally obtained with new physics at TeV scale. 	As interesting consequence, the mechanism may be directly probed  at the LHC by directly producing the  TeV states intrinsic of the mechanism. We show that the 3-3-1 model with right-handed neutrinos realizes naturally such seesaw mechanism.
%\\
%PACS: 14.60.St; 14.60.Pq; 12.60.Cn; 12.60.Fr.
%
\end{abstract}

\maketitle

\section{Introduction}
\label{sec:Introduction}
One of the major puzzles in  particle physics is the understanding of the small but nonzero neutrino masses established by the phenomenon of neutrino oscillation detected in solar\cite{solar} and atmospheric\cite{atm} as long as in laboratory experiments with reactors\cite{reactor} and acellerator\cite{acel} neutrinos. 

On the theoretical side, seesaw mechanisms is considered the most elegant way of generating small masses for the neutrinos. In conventional seesaw mechanisms\cite{typeI,typeII,typeIII}, neutrino masses get suppressed by a high-scale $M$ in its denominator according to the formula: $m_{\nu_l} = \frac{v^2_{ew}}{M}$ where $v_{ew}$ is a energy scale parameter at the electroweak range.

Regions of values for $M$ of interest in particle physics is $M$ close to unification scale($10^{12-14}$GeV)  and $M$ around TeV scale($1-10$TeV). In the first scenario, neutrino physics is associated to  theories of grand unification  and the quotient $\frac{v^2_{ew}}{M}$ lies already at the eV range, explaining in this way the smallness of the neutrino masses.  However, the unpleasant point of this scenario is that, due to the high value for $M$, such scenario can not be tested experimentally. 

In the second scenario, i.e., $M$ at the TeV range, for $v_{ew}$ at electroweak scale, we do not have, in fact, a seesaw mechanism because $v^2_{ew}$ is of order of  $M$ which leaves the quotient $\frac{v^2_{ew}}{M}$  close to the unity requiring, in this way,   huge fine tuning of the Yukawa couplings in order to generate neutrino mass at eV scale. People get interested in such scenario because it may be probed at LHC\cite{seesawTeV}. However, for we have a seesaw mechanism with $M$ at the TeV scale,  we need  $v_{ew}\leq 1$MeV. This is unnatural, unless we provide a supplementary mechanism of suppression for  $v_{ew}$. 

An interesting possibility for having a seesaw mechanism working at TeV scale is fitting the type II seesaw mechanism into the type I seesaw mechanism. For this we have to develop type II seesaw mechanism for other scalar multiplets than Higgs doublet. This idea was originally proposed by Ma at Ref. \cite{ma} and after generalized by Grimus {\it et al.} at Ref. \cite{grimus}.

 In this work we review this idea and show that it leads to a neutrino mass formula that  get suppressed by high-scale $M^3$ in its denominator. Due to this suppression factor, we call this mechanism ``triple seesaw mechanism''. We also show that the 3-3-1 model with right-handed neutrinos realizes naturally the triple seesaw mechanism. 

This works is organized as following. In Sec. (\ref{sec:AToyModel})  we make use of a toy model to present the idea. In Sec. (\ref{sec:ImplementingTheMechanismIntoTheStandardModel}) we develop the mechanism in an extension of the standard model. In Sec. (\ref{sec:ImplementingTheMechanismIntoThe331ModelWithRightHandedNeutrinos}) we implement the mechanism inside the 3-3-1 model, and finally in Sec. (\ref{sec:Summary}) we  present a summary of our results.
%%%%%%%%%%%%%%%%%%%%%%%%%%%%%%%%%%%%%%%%%%%%%%%%%%%%%%%%%%%
\section{A toy model}
\label{sec:AToyModel}

For presenting the idea we consider a simple toy model without reference at first  to any gauge model. The toy model involve one left-handed neutrino, $\nu_L$, one right-handed neutrino, $\nu_R$ and  two scalar fields $\phi_1$ and $\phi_2$. These scalars do not carry lepton number. We assume that the lagrangian  respects the set of discrete symmetry $(\nu_L, \nu_R, \phi_2)\rightarrow -(\nu_L, \nu_R, \phi_2)$. Besides, we allow that only $\nu_R$ develop Majorana mass term. After all this, the lagrangian we can build with these fields is formed by the terms
\begin{eqnarray}
	{\cal L}= -y\bar \nu_L \nu_R \phi_1 -M \bar \nu^C_R \nu_R + V(\phi_1,\phi_2).
	\label{toymodel}
\end{eqnarray}
When $\phi_1$ develop  vacuum expectation value(VEV) different from zero, $v_1$, the first term in the lagrangian above generates Dirac mass term for the neutrinos.  The first two terms of the lagrangian above provide the following  mass matrix for the neutrinos  in the basis $(\nu_L\,,\,\nu^C_R)$,
\begin{eqnarray}
	M_\nu =\left( \begin{array}{cc}
0 & yv_1 \\
yv_1 & M
\end{array}
\right).
\label{tomassmatrix}
\end{eqnarray}
Assuming  $M>> v_1$, we obtain after diagonalize this mass matrix
\begin{eqnarray}
	m_{\nu_l}\approx \frac{y^2v_1^2}{M}\,\,\,,\,\,\, m_{\nu_h}\approx M.
	\label{toyseesawI}
\end{eqnarray}
This is nothing more than the type I seesaw mechanism.

Now let us move to the scalar sector of the model. The main goal here is to  arrange things such that $v_1$ get suppressed by a factor $\frac{1}{M}$. For this we apply the type II seesaw mechanism to $\phi_1$.

The potential we can construct with all these scalar fields and respect the discrete symmetry mentioned above  is composed by the terms
\begin{eqnarray}
	V(\phi_i)&=&\mu^2_1 \phi^2_1+\mu^2_2 \phi^2_2+\lambda_1 \phi^4_1+\lambda_2 \phi^4_2+\lambda_3 \phi^2_1 \phi^2_2+\nonumber \\
	&&-\frac{M_1}{2}\phi_1 \phi_2^2 +\mbox{H.c}.
	\label{potential}
\end{eqnarray}
The condition for $V(\phi_i) $ develops minimum involves two constraint equations. However, what matter for us here is the one related to the VEV of $\phi_1$, 
\begin{eqnarray}
	v_1\big( \mu^2_1+\lambda_1 v^2_1+\frac{\lambda_3}{2}v^2_2\big) -M_1v^2_2=0,
	\label{minimumconditions1}
\end{eqnarray}
where $v_1$  and $v_2$ are the VEV's of $\phi_1$  and $\phi_2$, respectively.

The suppositions we take here are that $\phi_1$ is heavier than $\phi_2$, i.e., $\mu_1>> \mu_{2}$   and that $\mu_1\approx M_1$.  With this we find that the equation above provides,
\begin{eqnarray}
	v_1\approx \frac{v^2_2}{M_1}.
	\label{typeIIsstoy}
\end{eqnarray}

Considering that $M_1\approx M$ and substituting Eq. (\ref{typeIIsstoy})  in (\ref{toyseesawI}), we obtain,
\begin{eqnarray}
	m_{\nu_l}\approx \frac{y^2v^4_2}{M^3}.
	\label{toydoubless}
\end{eqnarray}
This is the neutrino mass formula  that arises in the triple seesaw mechanism. Perceive that there is a real possibility for we have neutrino masses at eV  range with new physics at  TeV scale. This is so because neutrino masses get suppressed by  high-scale $M^3$ in its denominator.  For example, for $M=(1-10)$TeV, we have a factor of suppression in the range $10^{9-12}$ which is sufficient to generate neutrino masses at around eV.

In the next two sections we implement the mechanism in the framework of realistic  neutrino models.
%%%%%%%%%%%%%%%%%%%%%%%%%%%%%%%%%%%%%%%%%%%%%%%%%%%%%%%%%%%%
\section{Implementing the mechanism into the standard model}
\label{sec:ImplementingTheMechanismIntoTheStandardModel}

The way we implement the mechanism here is different from the one done in the original work \cite{ma}. We  extend the standard model(SM) with some new fields. For sake of simplicity, we  consider only  one generation of fermions. The mechanism requires we add to the standard  field content one right-handed neutrino $\nu_R$ and  at least two scalar doublets $H_1$, $H_2$, plus a scalar singlet $S$. These scalars do not carry lepton number. To avoid undesirable terms, we assume that the whole lagrangian respects the discrete symmetry  $(H_1,S,\nu_R)\rightarrow -(H_1,S,\nu_R)$.  With this, the lagrangian  must involve the terms
\begin{eqnarray}
	{\cal L} \supset && -y\bar L \tilde{H_1} \nu_R - M\bar \nu^C_R \nu_R +V(H_1,H_2,S) 
	\label{SMdodified}
\end{eqnarray}
where $L=(\nu_L\,,\,e^-_L)^T$ is the standard lepton doublet. 

When $H_1$ develop nonzero VEV, $v_1$, the neutrinos gain Dirac mass term that together with the Majorana mass term for the right-handed neutrino form the following mass matrix in the basis $(\nu_L\,,\,\nu^C_R)$,
\begin{eqnarray}
	M_\nu =\left( \begin{array}{cc}
0 & yv_1 \\
yv_1 & M
\end{array}
\right).
\label{massmatrix}
\end{eqnarray}
Considering $M>> v_1$, we obtain after diagonalize this mass matrix,
\begin{eqnarray}
	m_{\nu_l}\approx \frac{y^2v_1^2}{M}\,\,\,,\,\,\, m_{\nu_h}\approx M.
	\label{SMseesawI}
\end{eqnarray}
Again, this is the type I seesaw mechanism.

Now let us move to the scalar sector of the model and apply the type II seesaw mechanism to the doublet $H_1$. The potential composed by this scalar content and that respect the set of discrete symmetry we discussed above is,
\begin{eqnarray}
	V(H_1,H_2,S)&=& \mu^2_1 (H_1^{\dagger} H_1)+\mu^2_2 (H_2^{\dagger} H_2)+\mu^2_SS^*S+\lambda_1 (H_1^{\dagger} H_1)^2+\lambda_2 (H_2^{\dagger} H_2)^2 \nonumber \\
	&&+\lambda_3 (S^* S)^2+\lambda_4 (H_1^{\dagger} H_1)(H_2^{\dagger} H_2)      +\lambda_5 (H_1^{\dagger} H_1)(S^* S) \nonumber \\
	&&+\lambda_6 (H_2^{\dagger} H_2)(S^* S)-\frac{M_1}{2}H_1^{\dagger} H_2 S +\mbox{H.c}.
\end{eqnarray}

The conditions for $V(H_1,H_2,S) $ develops minimum involve three constraint equations. What matter for us here is the  one related to the VEV of $H_1$,
\begin{eqnarray}
v_1\big( \mu^2_1+\lambda_1 v_1^22 +\frac{\lambda_4}{2}v^2_2 +\frac{\lambda_5}{2}v^2_S \big) -M_1v_2 v_S=0, 
	\label{minimumconditionSM}
\end{eqnarray}
where $v_2$  and $v_S$ are the VEV's of $H_2$  and $S$, respectively.

Let us suppose that  $H_1$ is  heavier than $H_2$ and $S$, i.e., $\mu_1>> \mu_2,\mu_S$, and  that the scale of energy, $M_1$, related to the trilinear term in the potential above is of the same order of magnitude of the mass of $H_1$, i.e., $\mu_1 \approx M_1$. With these two suppositions, the equation above provides,
\begin{eqnarray}
	v_1\approx \frac{v_2v_s}{M_1}.
	\label{typeIISSintheSM}
\end{eqnarray}
This is nothing more that the type II seesaw mechanism applied to $v_1$.

Now, considering $M_1 \approx M$ and combining Eq. (\ref{typeIISSintheSM}) with (\ref{SMseesawI}), we obtain,
\begin{eqnarray}
	m_{\nu_l}\approx \frac{y^2v^2_2v^2_s}{M^3}.
	\label{DoubleSSSM}
\end{eqnarray}
Note that, as expected, the mass of the light neutrino get suppressed by high-scale $M^3$ in its denominator. Moreover, perceive that the formula above involve two VEV's in its numerator. This gives us  more liberty to arrange the VEV's in conjunction with $M$ in order to obtain neutrino mass at eV range with $M$ at the  TeV scale.

For example, in this particular extension of the SM, $v_1$ develops exclusively Dirac mass terms for the neutrinos and $v_2$  get in charge of generating  masses for all standard charged fermions of the model. This requires $v_2\approx10^2$GeV. On the other hand,  $v_S$ is not constrained, which means it can take any value in the electroweak range. Being conservative and taking  $v_S=10^{-1}$GeV, $v_2 \approx 10^2$GeV  and  $M=10$TeV, we obtain
\begin{eqnarray}
	m_{\nu_l}\approx 0.1 y^2\mbox{eV}.
	\label{DoubleSSSMprediction}
\end{eqnarray}
We see that the mechanism provides  a real possibility for explaining solar and atmospherics neutrino oscillation with Yukawa couplings of order ${\cal O}(1)$.

From a phenomenological viewpoint, this scenario is interesting because it involves sizable couplings of the TeV new states with the standard leptons through the term $y\bar L \tilde{H_1} \nu_R$, allowing the mechanism  be directly probed  at the LHC by directly producing the  TeV states intrinsic of the mechanism.
\section{Implementing the mechanism into the 3-3-1 model with right-handed neutrinos}
\label{sec:ImplementingTheMechanismIntoThe331ModelWithRightHandedNeutrinos}

In this section we show that the triple seesaw mechanism takes place naturally in the gauge models based in the $SU(3)_C \otimes SU(3)_L \otimes U(1)_N$(3-3-1) symmetry called 3-3-1 model with right-handed neutrinos(331$\nu$R)\cite{footpp}.  In it the leptons come in triplet and singlets as follows,
\begin{eqnarray}
f_{aL} = \left (
\begin{array}{c}
\nu_a \\
e_a \\
\nu^{c}_a
\end{array}
\right )_L\sim(1\,,\,3\,,\,-1/3)\,,\,\,\,e_{aR}\,\sim(1,1,-1),
 \end{eqnarray}
with $a=1,2,3$ representing the three known generations. We are
indicating the transformation under 3-3-1 after the similarity
sign, ``$\sim$''. For the quark representation we refer the reader to the original work of the model\cite{footpp}

The  scalar content capable of generating masses for all charged fermions of the model is composed by three scalar triplets as follows\cite{footpp}
\begin{equation}
\chi=
\begin{pmatrix}
\chi^{0} \\
\chi^{-} \\
\chi'^{0}
\end{pmatrix}
,
\rho=
\begin{pmatrix}
\rho^{+} \\
\rho^{0} \\
\rho'^{+}
\end{pmatrix}
,
\eta=
\begin{pmatrix}
\eta^{0} \\
\eta^{-} \\
\eta'^{0}
\end{pmatrix}
,
\end{equation}
with $\eta$ and $\chi$ transforming as (1,3,$-\frac{1}{3}$) and $\rho$ transforming as (1,3,$-\frac{2}{3}$). Although the scalar content above involves  five neutral scalars, it is just necessary that $\eta^0\,,\,\rho^0$ and $\chi^{\prime 0}$ develop vacuum expectation value(VEV) for that all the particles of the model, with exception of the neutrinos, develop their correct mass terms. 

In the 331$\nu$R neutrino mass terms must, necessarily, involve the product  $\bar{f}_{L}f_{L}^{c}$. Considering its transformation  by the $SU(3)_L$ symmetry, $\bar{f}_{L}f_{L}^{c}=3^{*} \otimes 3^{*}=3 \oplus 6^{*}$, we see that, in order to generate mass terms for the neutrinos, we must couple  $\bar{f}_{L}f_{L}^{c}$ to an anti-triplet or to a sextet of scalars. Previous works showed that the first case leads to degenerate Dirac mass terms for the neutrinos\cite{footpp}. This case is not of interest for us here. In regard to the scalar sextet, recent works have employed it  to implement the type I and type II seesaw mechanisms into the model\cite{typeI331,typeII331}. 

A remarkable fact about this sextet is that  after the  $SU(3)_C \otimes SU(3)_L \otimes U(1)_N$ symmetry breaking to the $SU(3)_C \otimes SU(2)_L \otimes U(1)_Y$ one, the sextet,
\begin{equation}
S=
\dfrac{1}{\sqrt{2}}
\begin{pmatrix}
\Delta^{0} & \Delta^{-} & \Phi^{0} \\
\Delta^{-} & \Delta^{--} & \Phi^{-} \\
\Phi^{0} & \Phi^{-} & \sigma^{0}
\end{pmatrix}
\label{sextet}
\sim (1,6,\frac{-2}{3}),
\end{equation}
splits into a triplet plus a doublet and a singlet of scalars,
\begin{eqnarray}
S \rightarrow \Delta_{({\bf 1},{\bf 3},Y_\Delta)} +\Phi_{{({\bf 1},{\bf 
2},Y_{\Phi})}} + \sigma^0_{({\bf 1},{\bf 1},Y_{\sigma^0})},
\end{eqnarray}
where $Y$ are the hypercharges of the respective multiplets, with $Y_\Delta=-2$, $Y_{\Phi}=-1$ and $Y_{\sigma^0}=0$, and 
\begin{eqnarray}
\Delta=\frac{1}{\sqrt{2}}\left(\begin{array}{cc}
\Delta^{0} & \Delta^{-}  \\
\Delta^{-} & \Delta^{--} 
\end{array}\right )\,\,,\,\,\Phi=\frac{1}{\sqrt{2}}\left(\begin{array}{c}
\Phi^{0}   \\
 \Phi^{-} 
\end{array}\right )\,\,,\,\,\frac{\sigma^0}{\sqrt{2}}.
\label{breakofS} 
\end{eqnarray}
The Yukawa interaction $f^C_{L}S^*f_{L}$ can be dismembered  into the following ones, 
\begin{eqnarray}
\bar{f^C_{L}}S^*f_{L}  \rightarrow	\bar L^C \Delta^* L + \bar L \Phi \nu_R + \bar \nu^C_R \sigma_0 \nu_R,
\label{breakoftheyukawainteractions}
\end{eqnarray}
when the 3-3-1 symmetry breaks to the $SU(3)_c\times SU(2)_L\times U(1)_Y$ symmetry.
In the above expression $L$ represents the usual SM doublet of left-handed leptons.
It is easy to see that when $\Delta^0$, $\Phi^0$  and $\sigma^0$ develop nonzero VEV, then Dirac and Majorana mass terms are generated for the neutrinos. In this work we assume that 
$\Delta^0$ does not develop VEV. As we see, the model disposes of all  ingredients required by the triple seesaw mechanism.

When $\Phi^0$  and $\sigma^0$ develop nonzero VEV, let us suppose $v_\Phi$ and $v_\sigma$, the Yukawa interactions $h_{ab}\bar{f^C_{a_L}}Sf_{bL}$ generate the following mass terms for the neutrinos in the basis $(\nu_{e_L}\,,\,\nu_{\mu_L}\,,\,\nu_{\tau_L}\,,\,\nu^C_{e_R}\,,\,\nu^C_{\mu_R}\,,\,\nu^C_{\tau_R})^T=(\nu_L\,,\,\nu^C_R)^T$,
\begin{eqnarray}
\frac{1}{2}\left( \bar{\nu^C_{L}}\,\,,\,\, \overline{\nu_{R}}\right)
\left (
\begin{array}{cc}
0 & M_{D} \\
M_{D} & M_R
\end{array}
\right)
\left(
\begin{array}{c}
\nu_{L} \\
\nu^C_{R}
\end{array}\right),
\label{seesawmatrix}
\end{eqnarray}
where 
\begin{eqnarray}
	M_{D}=hv_{\Phi}\,\,\,\,\,\mbox{and}\,\,\,\,\,M_R=hv_{\sigma},
	\label{MDM}
\end{eqnarray}
with $h$  being a symmetric matrix formed by the Yukawa couplings $h_{ab}$. As it is very well known, from the diagonalization of the mass matrix in Eq.~(\ref{seesawmatrix}), for the case $v_{\sigma}>>v_{\Phi} $, we obtain, 
\begin{eqnarray}
m_{\nu_l} \simeq -M_{D}M_R^{-1}M_{D}, & m_h \simeq M_R,
\label{generalmassmatrix}
\end{eqnarray}
where $m_{\nu_l}$  is a mass matrix for the left-handed neutrinos and $m_h$  is the mass matrix for the right-handed neutrinos.  

On substituting the matrices $M_D$  and $M_R$ given in Eq.~(\ref{MDM}) into the Eq.~(\ref{generalmassmatrix}), we obtain,
\begin{eqnarray}
	m_{\nu_l}=-h\frac{v_\Phi^2}{v_\sigma}\,\,\,\,\,\mbox{and}\,\,\,\,\, m_h=hv_\sigma.
	\label{finalmass}
\end{eqnarray}
This is the type I seesaw mechanism that arises in the 331$\nu$R.

Let us move to the scalar sector. In order to avoid a proliferation of undesirable terms, we impose the potential respects the set of discrete symmetries $(\chi,\rho) \rightarrow -(\chi,\rho)$. After this, the most complete part of the potential that obeys this set of  discrete symmetry  and conserves lepton number is composed by the following terms,
\begin{eqnarray}
V& = & \mu_{\chi}^{2}\chi^{2}+\mu_{\eta}^{2}\eta^{2}+\mu_{\rho}^{2}\rho^{2}+\lambda_{1}\chi^{4}+\lambda_{2}\eta^{4}+\lambda_{3}\rho^{4}+\lambda_{4}(\chi^{\dagger}\chi)(\eta^{\dagger}\eta) \nonumber\\
  & + & \lambda_{5}(\chi^{\dagger}\chi)(\rho^{\dagger}\rho)+\lambda_{6}(\eta^{\dagger}\eta)(\rho^{\dagger}\rho)+\lambda_{7}(\chi^{\dagger}\eta)(\eta^{\dagger}\chi)+\lambda_{8}(\chi^{\dagger}\rho)(\rho^{\dagger}\chi) \nonumber\\
  & + & \lambda_{9}(\eta^{\dagger}\rho)(\rho^{\dagger}\eta)+(\frac{f}{\sqrt{2}}\epsilon^{ijk}\eta_{i}\rho_{j}\chi_{k}+H.C)+\mu_{S}^{2}Tr(S^{\dagger}S) \nonumber\\
  & + & \lambda_{10}Tr(S^{\dagger}S)^{2}+\lambda_{11}[Tr(S^{\dagger}S)]^{2}+(\lambda_{12}\eta^{\dagger}\eta+\lambda_{13}\rho^{\dagger}\rho+\lambda_{14}\chi^{\dagger}\chi)Tr(S^{\dagger}S) \nonumber\\
  & + & \lambda_{15}(\epsilon^{ijk}\epsilon^{lmn}\rho_{n}\rho_{k}S_{li}S_{mj}+H.C)+\lambda_{16}(\chi^{\dagger}S)(S^{\dagger}\chi)+\lambda_{17}(\eta^{\dagger}S)(S^{\dagger}\eta) \nonumber\\
  & + & \lambda_{18}(\rho^{\dagger}S)(S^{\dagger}\rho),
\end{eqnarray}
while the other part  that violates explicitly the lepton number is composed by the terms,
\begin{eqnarray}
V^{\prime} & = &
\lambda_{19}(\eta^{\dagger}\chi)(\eta^{\dagger}\chi)+\bigg[\frac{\lambda_{20}}{\sqrt{2}}\epsilon^{ijk}\eta_{m}^{*}S_{mi}\chi_{j}\rho_{k}+\frac{\lambda_{21}}{\sqrt{2}}\epsilon^{ijk}\chi_{m}^{*}S_{mi}\eta_{j}\rho_{k}\nonumber\\
        & - & M_{1}\eta^{T}S^{\dagger}\eta-M_{2}\chi^{T}S^{\dagger}\chi+H.C. \bigg],
\end{eqnarray}
where $M_1$ and $M_2$ are energy parameters associated to the explicit violation of the lepton number. 

We already assumed that only two of the three neutral scalars that compose the sextet will develop nonzero VEV. In regard to the triplets, they involve five neutral scalars. It was discussed above that, in order to generate the correct masses for all  charged fermions it is just necessary that only three of them develop nonzero VEV, namely $\eta^0\,,\,\rho^0$ and $\chi^{\prime 0}$. However, as we will see below, the seesaw mechanism we develop here requires that either $\eta^{\prime 0}$, or $\chi^{0}$ or both develop nonzero VEV. For sake of simplicity, we pick out $\eta^{\prime 0}$. Thus, the set of neutral scalars that will develop nonzero VEV is,
\begin{eqnarray}
\eta^{\prime 0}, \eta^0 , \rho^0 , \chi^{\prime 0}, \Phi^0 , \sigma^0 \rightarrow  \frac{1}{\sqrt{2}} (v_{   \eta^{\prime} ,\eta ,\rho ,\chi^{\prime}, \Phi , \sigma^0} 
+R_{\eta^{\prime} , \eta ,\rho ,\chi^{\prime}, \Phi^0 , \sigma^0} +iI_{\eta^{\prime} ,\eta ,\rho ,\chi^{\prime}, \Phi^0 , \sigma^0}). 
\label{setofvacua} 
\end{eqnarray}
Considering the shift of the neutral scalars above, the conditions for the minimum of the composite potential $V+V^{\prime}$ involve six equations. However, as the idea is to apply the type II seesaw mechanism to $\Phi^0$,  it is just necessary to analyze  the constraint equation related to the $v_\Phi$,
\begin{eqnarray}
	&& v_\Phi\bigg[ \mu_{S}^{2}+\frac{\lambda_{10}}{2}(v_{\Phi}^{2}+v_{\sigma}^{2})+\frac{\lambda_{11}}{2}(v_{\sigma}^{2}+2v_{\Phi}^{2})+\dfrac{\lambda_{12}}{2}(v_{\eta}^{2}+v_{\eta'}^{2})+\dfrac{\lambda_{13}}{2}v_{\rho}^{2}+\dfrac{\lambda_{14}}{2}v_{\chi'}^{2}-\lambda_{15}v_{\rho}^{2}+\nonumber\\
  &   & \dfrac{\lambda_{16}}{4}v_{\chi'}^{2}+\dfrac{\lambda_{17}}{4}\left( v_{\eta}^{2}+v_{\eta'}^{2}+\dfrac{v_{\eta}v_{\eta'}v_{\sigma}}{v_{\Phi}}\right)\bigg]-M_{1}v_{\eta}v_{\eta'}-\frac{\lambda_{20}+\lambda_{21}}{4}(v_{\eta'}v_{\rho}v_{\chi'})=0.
  \label{minimumoverphi}
\end{eqnarray}

All the massive components of the sextet $S$ and the triplet $\chi$ gain mass when the 3-3-1 symmetry breaks to the electroweak symmetry. The order of magnitude of such masses lie around the 331 symmetry breaking scale. In view of this we assume $\mu_S\approx \mu_\chi\equiv M$. Besides, we consider that the  energy scale associated to the explicit violation of the lepton number, $M_1 \,,\, M_2$, also lies around the 331 symmetry breaking scale, i.e., $M_1\approx M_2 \equiv M$. After such considerations, the equation above  provides the relation,
\begin{eqnarray}
	v_\Phi \cong \frac{v_\eta v_{\eta^{\prime}}}{M}.
	\label{typeII331}
\end{eqnarray}
Now we came to the main point of this section. On substituting Eq. (\ref{typeII331})  in Eq. (\ref{finalmass}), we obtain the following formula to the  neutrino masses,
\begin{eqnarray}
	m_{\nu_l}=-h\frac{v^2_\eta v^2_{\eta^{\prime}}}{M^3}.
	\label{finalmassII}
\end{eqnarray}
As showed, neutrino masses get suppressed by high-scale $M^3$ in its denominator. This allows we have neutrinos with masses  in the eV range with $M$ around few TeV's. 

Let us discuss a little the values of the parameters involved in the triple seesaw mechanism neutrino mass formula above. It is expected in the model that $ v^2_\rho + v^2_\eta \approx (246\mbox{GeV})^2$. On the other hand, the main consequence of a nonzero VEV for $\eta^{\prime 0 }$ is the rising of a mixing  among the ordinary quarks with the exotic ones. In view of this, it is natural to expected a low value for  $v_{\eta^{\prime}}$. In fact, the upper bound $v_{\eta^{\prime}}< 40$GeV  was obtained in Ref. \cite{bound}.  Thus, on being conservative, and taking  $v_\eta \approx 10^2$GeV, $v_{\eta^{\prime}}\approx10^{-1}$GeV  and $M=10$TeV, we obtain
\begin{eqnarray}
	m_{\nu_l}=-0.1h \mbox{eV}.
	\label{prediction}
\end{eqnarray}
This requires also little fine tuning of the Yukawa couplings $h$'s to generate  neutrino masses small enough in order to explain solar and atmospherics neutrino oscillation experiments. Thus the mechanism can be verified at the LHC through the direct detection of the TeV states of the 331$\nu$R involved in the mechanism.

We say that the triple seesaw mechanism is a natural outcome of the 331$\nu$R for two basic reasons. First, the high-scale  $M$ coincides with the 331 symmetry breaking  scale , and second, the VEV $v_{\eta^{\prime}}$ that appears in the numerator of the neutrino mass formula is phenomenologically small. 
%%%%%%%%%%%%%%%%%%%%%%%%%%%%%%
\section{Summary}
\label{sec:Summary}

In this work we developed a new kind of seesaw mechanism  that is really capable of explaining the smallness of neutrino masses  with new physics just at TeV scale. It arises from the fitting of the type II seesaw mechanism into the type I seesaw mechanism.   We first presented the idea through a toy model built specially to accommodate the mechanism. As realistic scenario, we implemented the mechanism in an extension of the SM involving right-handed neutrinos and some new scalar fields. Next we developed the mechanism in the framework of the 331$\nu$R. The main result of such fitting in both models is that   neutrino masses formula get suppressed by the factor  $M^3$ in its denominator. Thus it is easy to see that for $M=1-10$TeV, neutrino masses get suppressed by the factor $10^{9-12}$. This is sufficient to guarantee that neutrino will gain masses at eV scale.  Due to this suppression factor, we call this mechanism ``triple seesaw mechanism''. However, we would like to emphazise that the mechanism is a natural outcome of the 331$\nu$R. From phenomenological viewpoint, the interesting aspect of such mechanism is that its signature may be probed at LHC through direct detection of their intrinsic TeV states. 
%%%%%%%%%%%%%%%%%%%%%%%%%%%%%%%%%%%%%%%%%%%%%%%%%%%%%%%%%%%%%%%%%%%%%%%%%%%%%%%%%%%%%
\acknowledgments
%%%%%%%%%%%%%%%%%%%%%%%%%%%%%%%%%%%%%%%%%%%
 This work was supported by Conselho Nacional de Pesquisa e
Desenvolvimento Cient\'{i}fico- CNPq and Coordena\c c\~ao de Aperfei\c coamento de Pessoal de N\'{i}vel Superior - capes(PNPD/PROCAD).

%%%%%%%%%%%%%%%%%%%%%%%%%%%%%%%%%%%%%%%%%%%%%%%%%%%%%%%%%%%%%%%%%%%%%%%%%%%%%%%%%%%%%%%%%%%

%%%%%%%%%%%%%%%%%%%%%%%%%%%%%%%%%%%%%%%%%%%%%%%%%%%%%%%%%%%%%%%%
%%%%%%%%%%%%%%%%%%%%%%%%%%%%%%%%%%%%%%%%%%%%%%%%%%%%%%%%%%%%%%%%%%%%%%%%%%%%%%%%%%%%%%%%

\end{document}